\documentclass[prd,aps,preprint,showpacs,superscriptaddress]{revtex4}

\usepackage{latexsym,graphicx,epsfig,color,xspace}
\usepackage{amssymb,amsmath,amsxtra,amsfonts}

\usepackage{titlesec}

\titleformat{\section}{\large\bfseries}{\thesection}{1em}{}

\newcommand{\Ds}{\ensuremath{D_s^+}\xspace}
\newcommand{\Bc}{\ensuremath{B_c^+}\xspace}
\newcommand{\DsStar}{\ensuremath{D_s^{*+}}\xspace}

\newcommand{\R}{\ensuremath{\mathcal{R}}}

\newcommand*{\Jpsi}{\ensuremath{J/\psi}\xspace}
\newcommand{\Br}{\ensuremath{\mathcal{B}}}

\newcommand{\bea}{\begin{eqnarray}}
\newcommand{\ena}{\end{eqnarray}}
\newcommand{\be}{\begin{equation}}
\newcommand{\en}{\end{equation}}
\newcommand{\nn}{\nonumber\\}
\newcommand{\ed}{\end{document}} 

\newcommand{\ord}{\mathcal{O}}

\begin{document}

%\title{Semileptonic and nonleptonic decays of the $B_c$ meson revisited}
\title{Study of $B_c$ decays into charmonia and $D$ mesons}

\author{S.~Dubni\v{c}ka}
\affiliation{Institute of Physics, Slovak Academy of Sciences, 
Bratislava, Slovakia} 

\author{A.Z.~Dubni\v{c}kov\'{a} }
\affiliation{Comenius University, Bratislava, Slovakia}

\author{A.~Issadykov}
\email{isadykov@theor.jinr.ru}
\affiliation{
Joint Institute for Nuclear Research, Dubna, Russia}

\author{M.A.~Ivanov}
\email{ivanovm@theor.jinr.ru}
\affiliation{
Joint Institute for Nuclear Research, Dubna, Russia}

\author{A.~Liptaj}
\email{andrej.liptaj@savba.sk}
\affiliation{Institute of Physics, Slovak Academy of Sciences, 
Bratislava, Slovakia} 

\begin{abstract}
In the wake of recent measurements of the decays
$B_c^+\to J/\psi D^+_s$ and $B_c^+\to J/\psi D^{\ast\,+}_s$
performed by the LHCb and ATLAS Collaborations,
we recalculate their branching fractions in the framework
of the covariant confined quark model. 
We compare the obtained results with available experimental
data, our previous findings and numbers from other approaches.  
\end{abstract}


\maketitle

\section{Introduction}
\label{sec:intro}

Recently the ATLAS Collaboration reported on the measurement
of the various branching fractions of the decays
$B_c^+ \to J/\psi D_s^+$ and $B_c^+ \to J/\psi D_s^{*+}$ \cite{Aad:2015eza}.
The first observations of these decays have been performed by
the LHCb Collaboration \cite{Aaij:2013gia}. In view of these developments,
we decided to recalculate the amplitudes and branching fractions within
the covariant confined quark model. Our previous study of exclusive semileptonic
and nonleptonic decays of the $B_c$ meson was done more than ten years ago
within a relativistic constituent quark model 
\cite{Ivanov:2006ni,Ivanov:2005fd,Ivanov:2002un,Ivanov:2000aj}.
The modern approach with embedded infrared confinement
[for short, covariant confined quark model (CCQM)],
is a successor of the previous approach.  
Due to the confinement feature it has  more wide region of applications.

Many facets of the $B_c$ production were discussed in theoretical papers
by Likhoded and his co-authors, see, e.g. 
\cite{Berezhnoy:1995au,Berezhnoy:1996ks,Berezhnoy:1997fp}.
The decay properties of the above processes were studied in various theoretical
approaches \cite{Chang:1992pt,Colangelo:1999zn,Kiselev:2002vz,Kiselev:2003mp,Ebert:2003cn,Likhoded:2009ib,Dhir:2008hh,Ke:2013yka,Rui:2014tpa,Kar:2013fna,Zhu:2017lqu}. 
The decays $B_c^+ \to J/\psi D_s^+$ and $B_c^+ \to J/\psi D_s^{*+}$ proceed
via $b\to c\bar c s$ transition which is theoretically described
by the effective Hamiltonian with the relevant Wilson coefficients.
The physical amplitudes are described by 
color-enhanced, color-suppressed and annihilation diagrams. 
The two first diagrams are factorized into the leptonic decay part
and the transition of the $B_c$ meson into charmonium or
$D$ meson. The theoretical description of this transition gives
the most sizable uncertainties to the predicted physical observables.

In the paper \cite{Chang:1992pt} Heavy Quark Effective Theory (HQET)
in combination with suitable Bethe-Salpeter kernel was used to evaluate
the form factors. 
The form factors were computed in \cite{Colangelo:1999zn} as overlap integral 
of the meson wave-functions obtained using a QCD relativistic potential model. 
In the papers \cite{Kiselev:2002vz,Kiselev:2003mp} the $B_c$ decays
have been studied in the framework of QCD sum rules.
Semileptonic and nonleptonic decays of the $B_c$ meson to charmonium and a $D$ 
meson were studied in the framework of the relativistic quark model in
\cite{Ebert:2003cn}. The decay form factors were expressed through the overlap 
integrals of the meson wave functions in the whole accessible kinematical 
range. 
Decays $B_c \to J/\psi + n\pi$ were considered in \cite{Likhoded:2009ib}. 
Using existing parametrizations for $B_c\to J/\psi$ form-factors and 
$W\to n\pi$ spectral functions,  branching fractions and transferred momentum 
distributions have been calculated. 
An analysis of the $B_c$ form factors in the Wirbel-Stech-Bauer (WSB) framework
has been performed in \cite{Dhir:2008hh}. Branching ratios of two body 
decays of $B_c$ meson to pseudoscalar and vector mesons were obtained. 
In the paper \cite{Ke:2013yka}  form factors for the transitions 
$B_c\to J/\psi$ and $B_c\to \psi(2S)$ have been calculated
within the light-front quark model (LFQM) numerically. 
Then the partial widths of the semileptonic and nonleptonic decays
have been determined. 
A systematic investigation of the two-body nonleptonic decays 
$B_c\to J/\psi (\eta_C) + P(V)$ was performed in \cite{Rui:2014tpa} 
by employing the perturbative QCD approach based on the $k_T$ factorization.
The exclusive nonleptonic $B_c\to VV$ decays were studied in \cite{Kar:2013fna}
within the factorization approximation, in the framework of the relativistic 
independent quark model, based on a confining potential in the scalar-vector 
harmonic form. 
In the recent paper \cite{Zhu:2017lqu} the form factors of 
the transition of $B_c$ meson into $S$-wave charmonium 
were  investigated within  the nonrelativistic QCD effective theory. 
The next-to-leading order relativistic corrections to the form factors
were obtained.

\section{Effective Hamiltonian and matrix element}
\label{sec:EffHam}

The effective Hamiltonian describing the $B_c$ nonleptonic
decays into charmonium and $D(D_s)$ meson is given by 
(see, Ref.~\cite{Buchalla:1995vs})
\bea
{\mathcal H}_{\rm eff} &=&
-\frac{G_F}{\sqrt{2}}\,V_{cb} V^\dagger_{cq}\,\sum_{i=1}^6 C_i\,\ord_i,
\nn
&&\nn
\ord_1 &=& (\bar c_{a_1} b_{a_2})_{V-A} (\bar q_{a_2} c_{a_1})_{V-A}, 
\qquad 
\ord_2  =  (\bar c_{a_1}\,b_{a_1})_{V-A}, (\bar q_{a_2}\,c_{a_2})_{V-A},
\nn
\ord_3 &=& (\bar q_{a_1} b_{a_1})_{V-A} (\bar c_{a_2} c_{a_2})_{V-A}, 
\qquad 
\ord_4  =  (\bar q_{a_1} b_{a_2})_{V-A} (\bar c_{a_2} c_{a_1})_{V-A},
\nn
\ord_5 &=& (\bar q_{a_1} b_{a_1})_{V-A} (\bar c_{a_2} c_{a_2})_{V+A}, 
\qquad 
\ord_6  =  (\bar q_{a_1} b_{a_2})_{V-A} (\bar c_{a_2} c_{a_1})_{V+A},
\label{eq:EffHam}
\ena
where the subscript $V-A$ refers to the usual left--chiral current
$O^\mu=\gamma^\mu(1-\gamma^5)$ and  $V+A$  to the usual right--chiral one
$O^\mu_+=\gamma^\mu(1+\gamma^5)$. The  $a_i$ denote the color indices.
The quark $q$ stands for either $s$ or $d$. 

The numerical values of the Wilson coefficients are taken from 
Ref.~\cite{Descotes-Genon:2013vna}. 
They were computed at the matching scale $\mu_0=2 M_W$ 
at the NNLO precision and run down to  the hadronic scale $\mu_b= 4.8$~GeV.
They are listed in Table~\ref{tab:WC}.
\bgroup 
\def\arraystretch{1.5}
\begin{table}[htbp]
\caption{Values of the Wilson coefficients.}
\vspace*{2mm}  
\centering
 \begin{tabular}{cccccc}
\hline
  $C_1$    &  $C_2$    &  $C_3$     &  $C_4$    & $C_5$  &  $C_6$   \\
\hline
 $-0.2632$ & $1.0111$ &  $-0.0055$ & $-0.0806$ & 0.0004 & 0.0009   \\
\hline
 \end{tabular}
\label{tab:WC}
\end{table}
\egroup
Since the numerical values of the  $C_5$  and  $C_6$ are negligibly small,
we drop the contribution from those operators.

\begin{figure}[ht]
\label{tab:leptonic} 
\begin{center}
\includegraphics[width=0.8\textwidth]{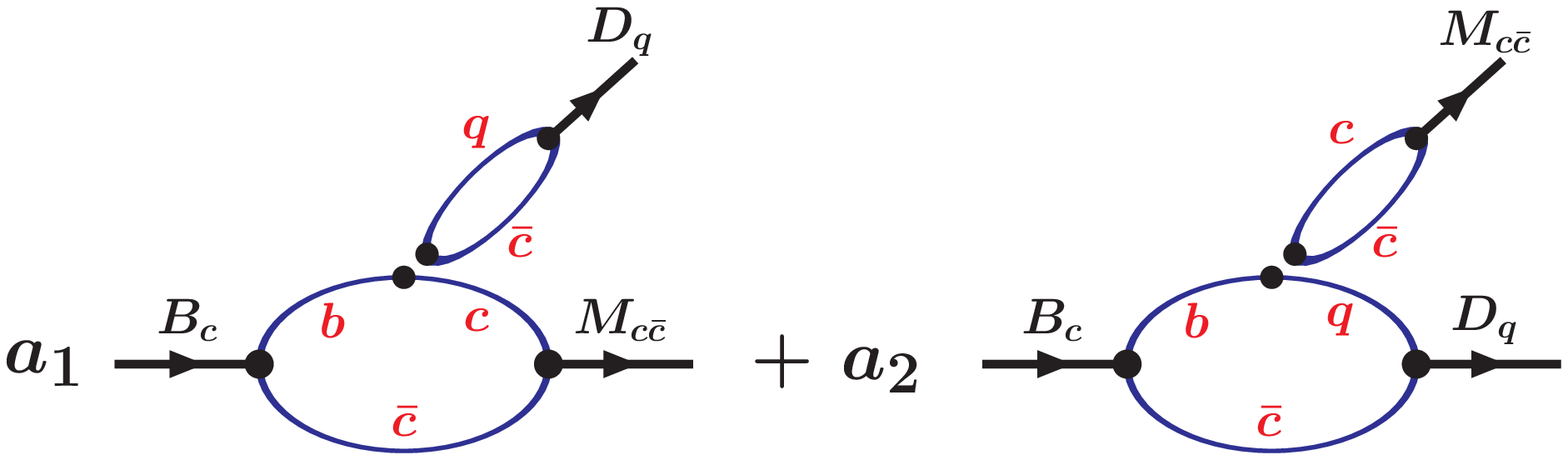} 
\end{center}
\caption{\label{fig:decay}
Pictorial representation of the matrix elements of  the nonleptonic 
$B_c$ decays.}
\end{figure}

By using the Fierz transformation one can check that $\ord_3=\ord_1$
and $\ord_4=\ord_2$. Then the calculation of the matrix elements
describing the nonleptonic decays of the $B_c$ meson  into charmonium and 
$D(D_s)$ meson is straightforward. Pictorial representation of the matrix 
elements is shown in Fig.~\ref{fig:decay}. 

The combinations of the Wilson coefficients appear as
$a_1=C_2+C_4 + \xi\,(C_1+C_3)$ and $a_2=C_1+C_3 +\xi\,(C_2+C_4)$
with $\xi=1/N_c$. 
In the numerical calculations we set the color-suppressed parameter $\xi$ 
to zero. Then the Wilson coefficients are equal to 
\be
a_1=C_2+C_4=0.93\,,\qquad\text{and}\qquad a_2=C_1+C_3=-0.27
\label{eq:WC-numer}
\en
 which should be compared with the old ones
 $a_1=1.14$ and $a_2=-0.20$ used in our previous paper \cite{Ivanov:2006ni}.

 One has to note that the signs in front of the leptonic decay constants $f_D$ 
and $f_{\eta_c}$ should be opposite to those defined in their leptonic decays.
It comes from the observation that the meson momentum flows
in the opposite direction  in the case of the nonleptonic decays
as compared with the case of the leptonic decays. 

\section{Invariant and helicity amplitudes}
\label{sec:helicity}

The invariant form factors for the semileptonic $B_c$ decay
into the hadron with spin $S=0,1$ are defined by

\bea
{\mathcal M^\mu_{\,S=0}} &=&
P^\mu\,F_+(q^2)+q^\mu\,F_-(q^2),
\label{eq:def-PP}\\[1.5ex]
{\mathcal M^\mu_{\,S=1}}&=&
\frac{1}{m_1+m_2}\,\epsilon^\dagger_\nu\,
\left\{\,
-\,g^{\mu\nu}\,Pq\,A_0(q^2)+P^\mu\,P^\nu\,A_+(q^2)+q^\mu\,P^\nu\,A_-(q^2)
\right.\nn
&&
\left.\hspace*{2.5cm}
+\,i\,\varepsilon^{\mu\nu\alpha\beta}\,P_\alpha\,q_\beta\,V(q^2)\right\},
\label{eq:def-PV}
\ena
where $P=p_1+p_2$ and $q=p_1-p_2$. Here $p_1$ is the momentum
of the ingoing meson with a mass $m_1$ ($B_c$) and
$p_2$ is the momentum of the outgoing meson with a mass $m_2$.
It is convenient to express all physical observables
through the helicity form factors $H_m$.
The helicity form factors $H_m$ can be written in terms of
the invariant form factors in the following way \cite{Ivanov:2000aj}:

\noindent Spin S=0:
\bea
H_t &=& \frac{1}{\sqrt{q^2}}
\left\{(m_1^2-m_2^2)\, F_+ + q^2\, F_- \right\}\,, \qquad
H_\pm = 0\,, \qquad
H_0 = \frac{2\,m_1\,|{\bf p_2}|}{\sqrt{q^2}} \,F_+ \,.
\label{eq:helS0}
\ena

\noindent Spin S=1:
\bea
H_t &=&
\frac{1}{m_1+m_2}\frac{m_1\,|{\bf p_2}|}{m_2\sqrt{q^2}}
\left\{ (m_1^2-m_2^2)\,(A_+ - A_0)+q^2 A_- \right\},
\nn
H_\pm &=&
\frac{1}{m_1+m_2}\left\{- (m_1^2-m_2^2)\, A_0
\pm 2\,m_1\,|{\bf p_2}|\, V \right\},
\label{eq:helS1}\\
H_0 &=&
\frac{1}{m_1+m_2}\frac{1}{2\,m_2\sqrt{q^2}}
\left\{-(m_1^2-m_2^2) \,(m_1^2-m_2^2-q^2)\, A_0
+4\,m_1^2\,|{\bf p_2}|^2\, A_+\right\}.
\nonumber
\ena
Here $|{\bf p_2}|=\lambda^{1/2}(m_1^2,m_2^2,q^2)/(2\,m_1)$
is the momentum of the outgoing meson
in the $B_c$ rest frame.

The nonleptonic $B_c$ decay widths in terms of the helicity
amplitudes are given by

\bea
\Gamma(B_c\to \eta_c D_q ) & = & 
N_W\,
\left\{
a_1 f_{D^-_q} m_{D^-_q}  H_t^{B_c\to \eta_c}(m^2_{D^-_q})
+
a_2 f_{\eta_c}m_{\eta_c} H_t^{B_c\to D^-_q}(m^2_{\eta_c})
\right\}^2\,,
\nn[1.2ex]
\Gamma(B_c\to \eta_c D_q^{\ast} ) & = &
N_W\,
\left\{
 a_1 f_{D^{\ast\,-}_q} m_{D^{\ast\,-}_q} H_0^{B_c\to \eta_c}(m^2_{D^{\ast\,-}_q})
-
 a_2 f_{\eta_c} m_{\eta_c} H_t^{B_c\to D^{\ast\,-}_q}(m^2_{\eta_c})
\right\}^2\,,
\nn[1.2ex]
\Gamma(B_c\to J/\psi D_q ) & = &
N_W\,
\left\{
- a_1 f_{D^-_q} m_{D^-_q} H_t^{B_c\to J/\psi }(m^2_{D^-_q})
+
  a_2 f_{J/\psi}m_{J/\psi} H_0^{B_c\to D^{-}_q}(m^2_{J/\psi})
\right\}^2\,,
\nn[1.2ex]
\Gamma(B_c\to J/\psi D_q^{\ast} )) & = &
N_W\,
\sum\limits_{i=0,\pm}
\left\{
 a_1 f_{D^{\ast\,-}} m_{D_q^{\ast\,-}} H_i^{B_c\to J/\psi}(m^2_{D_q^{\ast\,-}})
+
 a_2 f_{J/\psi} m_{J/\psi} H_i^{B_c\to D_q^{\ast\,-}}(m^2_{J/\psi})
\right\}^2\,,
\nonumber
\ena
where we use the short notation
\[
N_W \equiv \frac{G^2_F}{16\,\pi} \frac{ |{\mathbf p_2}|}{m^2_1}
|V_{cb}V_{cq}^\dagger |^2\,.
\]

\section{Form factors}

We calculate the relevant hadronic form factors in the framework
of the covariant confined quark model \cite{Branz:2009cd}.

The starting point of the CCQM is the effective Lagrangian describing
coupling of the given hadron with its interpolating quark current.
In particular, the coupling of a meson $M$ to its constituent quarks $q_1$ 
and $\bar q_2$ is given by the Lagrangian
\bea
{\cal L}_{\rm int}(x) &=& g_M\,M(x)\cdot J_M (x) + {\rm H.c.},
\nn
J_M (x) &=& \int\!\! dx_1\!\! \int\!\! dx_2\, F_M (x;x_1,x_2)\, 
\bar{q}_2 (x_2)\,\Gamma_M\, q_1(x_1),
\label{eq:lag}
\ena
where $g_M$ denotes the coupling strength of the meson with its constituent 
quarks, the Dirac matrix $\Gamma_M$ projects onto the relevant meson state, 
i.e., $\Gamma_M=I$ for a scalar meson, $\Gamma_M=\gamma^5$ for a 
pseudoscalar meson, and $\Gamma_M=\gamma^{\mu}$ for a vector meson. 
The vertex function $F_M$ is chosen in the translational invariant form 
\bea
F_M (x;x_1,x_2) &=& \delta(x - w_1 x_1 - w_2 x_2) \Phi_M((x_1-x_2)^2),
\nn
\Phi_M((x_1-x_2)^2) &=& \int\frac{d^4\ell}{(2\pi)^4} e^{-i\ell(x_1-x_2)}
\widetilde \Phi_M(-\ell^2), \quad\text{where}\quad 
\widetilde \Phi_M(-\ell^2) = e^{\ell^2/\Lambda^2_M}.
\label{eq:vertex}
\ena
Here $w_i = m_{q_i}/(m_{q_1}+m_{q_2})$ so that $w_1+w_2=1$, and
the parameter $\Lambda_M$ characterizes the meson size.
The matrix elements of the physical processes
are defined by the appropriate $S$-matrix elements with
the $S$-matrix being constructed by using the interaction Lagrangian
given by Eq.~(\ref{eq:lag}).  The $S$-matrix elements in the momentum
space are described by a set of Feynman diagrams which are presented 
as convolution of quark propagators and vertex functions.
The free local fermion propagator is used for  the constituent quark:
\be
\label{eq:prop}
S_q(k) = \frac{1}{ m_q-\not\! k -i\epsilon } = 
\frac{m_q + \not\! k}{m^2_q - k^2  -i\epsilon }
\en 
with an effective constituent quark mass $m_q$.  The coupling strength $g_M$ 
is determined by the so-called compositeness condition which was discussed
in our previous paper in great details, see, e.g. 
Refs.~(\cite{Efimov:1988yd,Efimov:zg,Branz:2009cd}).
The  infrared cutoff parameter $\lambda$ is introduced on the last step
of calculations which  effectively guarantees the confinement of quarks 
within hadrons. This method is quite general and can be used for diagrams with 
an arbitrary number of loops and propagators. 
In the CCQM the infrared cutoff parameter $\lambda$ is taken to be universal 
for all physical processes.

The model parameters are determined by fitting calculated quantities 
of basic processes to available experimental data or 
lattice simulations. In this paper we will use the updated least-squares fit
performed in Refs.~\cite{Gutsche:2015mxa,Ganbold:2014pua,Issadykov:2015iba}.
All necessary details of the calculations of the leptonic decay
constants and hadronic form factors may be found in our recent
publications \cite{Ivanov:2015tru,Ivanov:2017mrj,Ivanov:2015woa}.

The fitted values  of the meson size parameters are given in 
Table~\ref{eq:size-param}.
\bgroup 
%\begin{table}[htbp]
\begin{table}[htp]
\caption{Values of the meson size parameters in GeV.}
\vspace*{2mm}  
\centering
\def\arraystretch{1.5}
 \begin{tabular}{ccccccc}
\hline
$ \Lambda_{B_c}$ & $ \Lambda_{\eta_c} $ & $\Lambda_{J/\psi}$ & $\Lambda_D$ &
$ \Lambda_{D^\ast} $ & $\Lambda_{D_s}$ &  $\Lambda_{D^{\ast}_{s}}$  \\
\hline
  \ \ 2.73 \ \  & \ \ 3.87 \ \ & \ \ 1.74 \ \ &  \ \ 1.6\ \ 	& 
\ \ 1.53 \ \    & \ \ 1.75 \ \  & \ \ 1.56 \ \ \\ 
\hline
\end{tabular}
\label{eq:size-param}
\end{table}
\egroup

The calculated values of leptonic decay constants are given in 
Table~\ref{eq:lept}. Note that the decay constant $f_{\eta_c}$ was calculated by
using the size parameter $\Lambda_{\eta_c}$ which was obtained from 
fitting the branching ratio of the $\eta_c$ meson two-photon decay  
to its experimental value given in PDG~\cite{Olive:2016xmw}.
\bgroup 
%\begin{table}[htbp]
\begin{table}[htp]
\caption{The calculated values of leptonic decay constants in MeV.}
\vspace*{2mm}  
\centering
\def\arraystretch{1.5}
 \begin{tabular}{ccccccc}
\hline
$ f_{B_c}$ & $ f_{\eta_c} $ & $f_{J/\psi}$ & $f_D$ &
$ f_{D^\ast} $ & $f_{D_s}$ &  $f_{D^{\ast}_{s}}$  \\
\hline
  \ \ 489 \ \  & \ \ 628 \ \  & \ \ 415 \ \ &  \ \ 206\ \ 	& 
  \ \ 244 \ \  & \ \ 257 \ \  & \ \ 272 \ \ \\ 
\hline
\end{tabular}
\label{eq:lept}
\end{table}
\egroup

\begin{figure}[ht]
\begin{center}
\begin{tabular}{lr}
\includegraphics[width=0.40\textwidth]{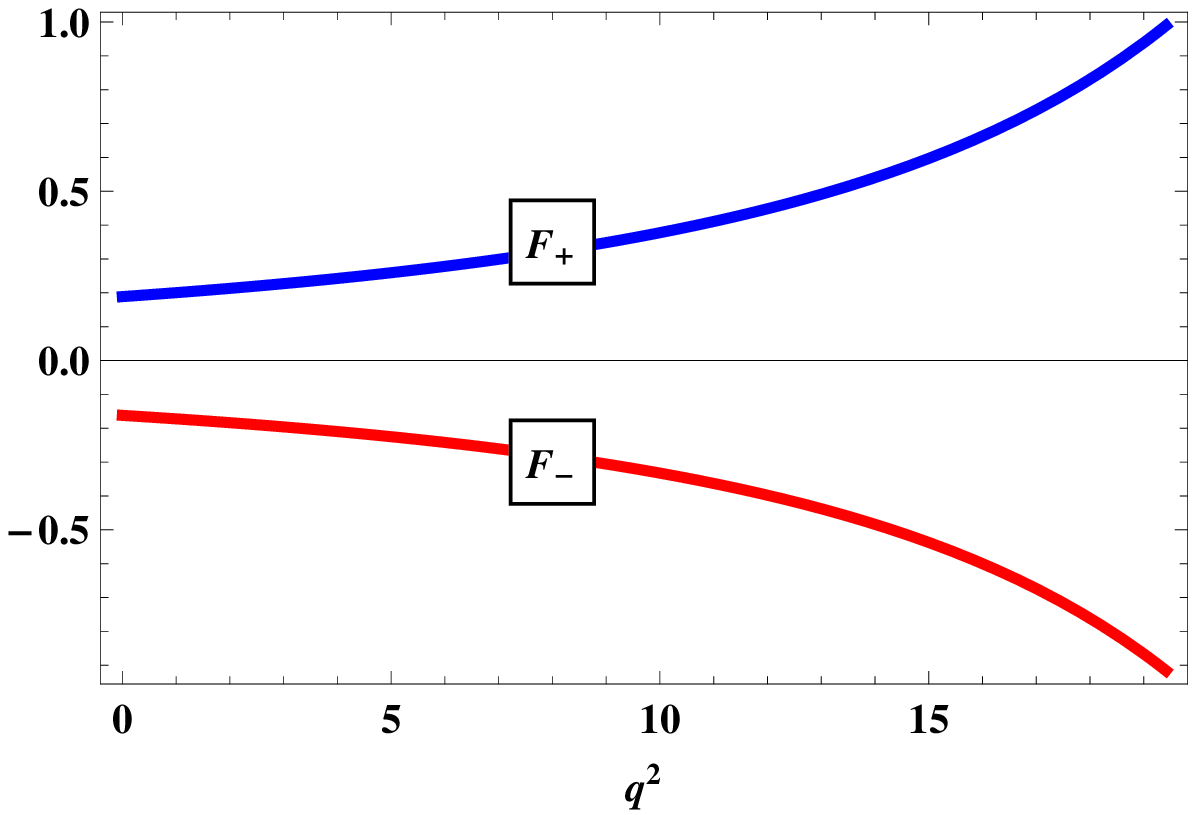} \qquad & \qquad
\includegraphics[width=0.40\textwidth]{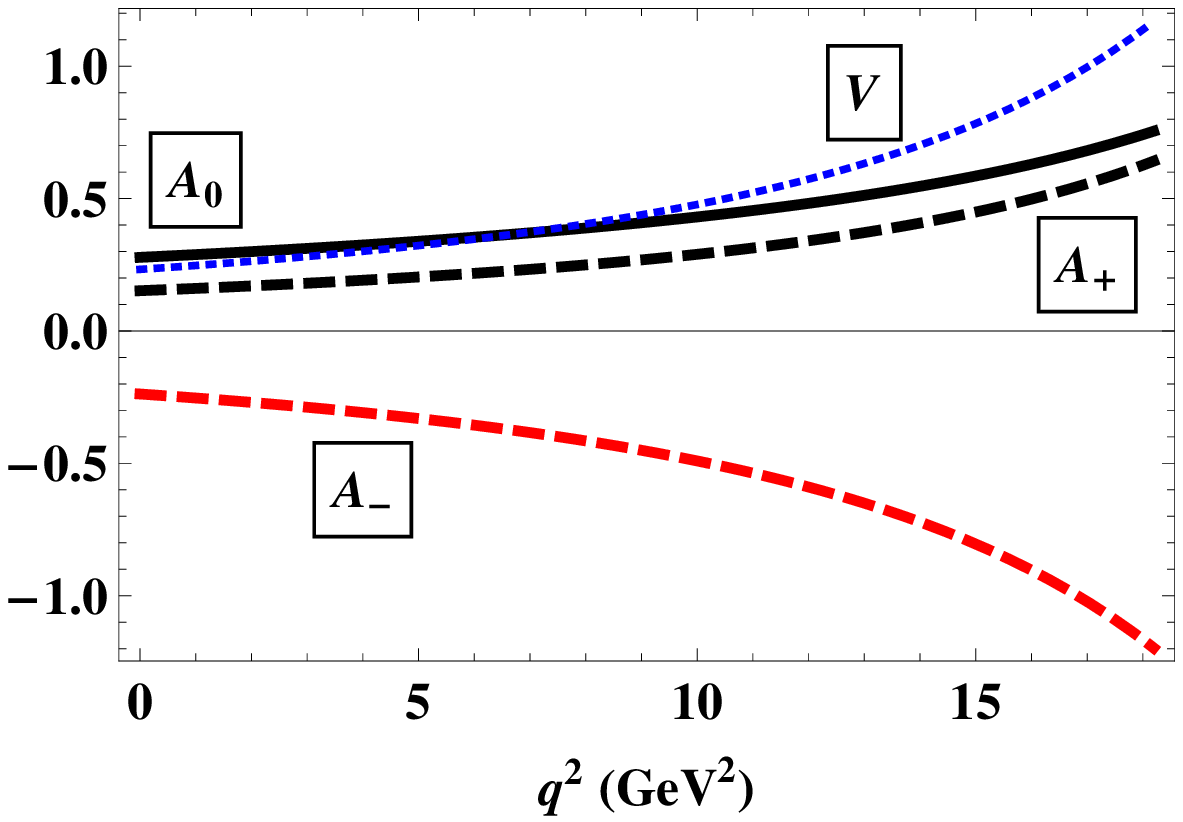} \\
\includegraphics[width=0.40\textwidth]{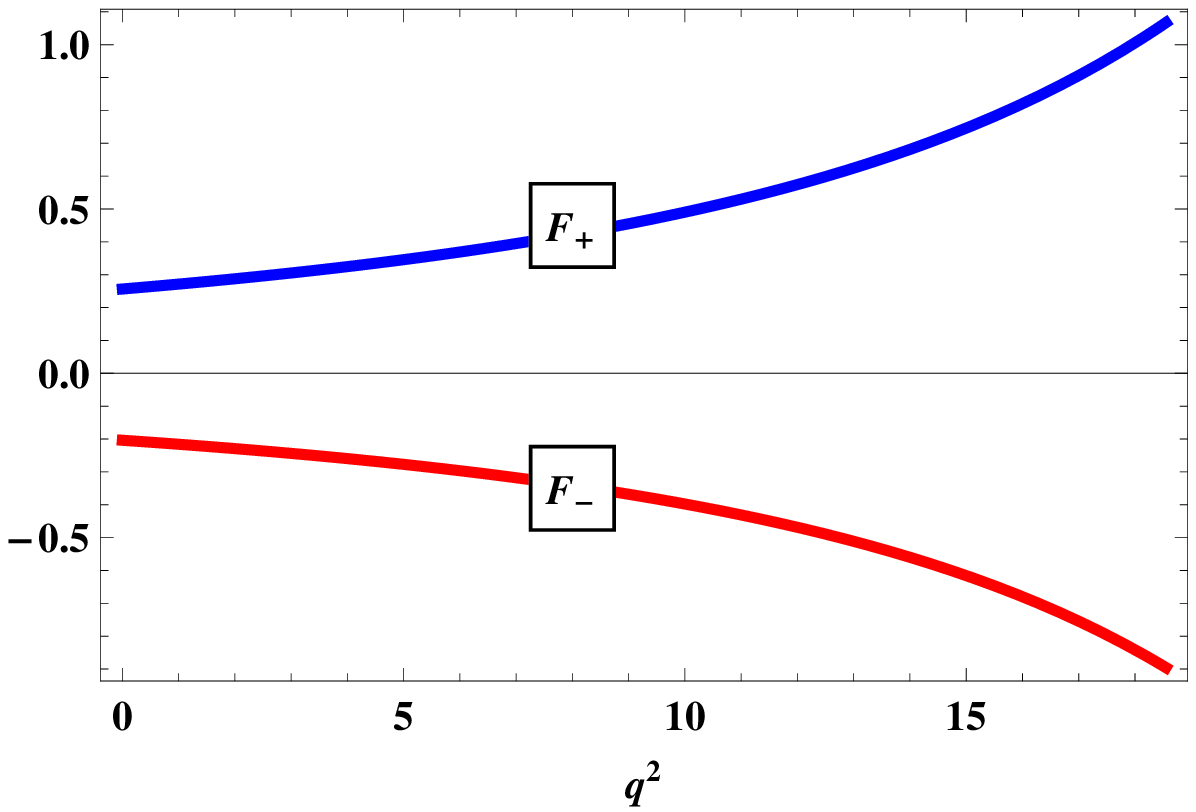} & 
\includegraphics[width=0.40\textwidth]{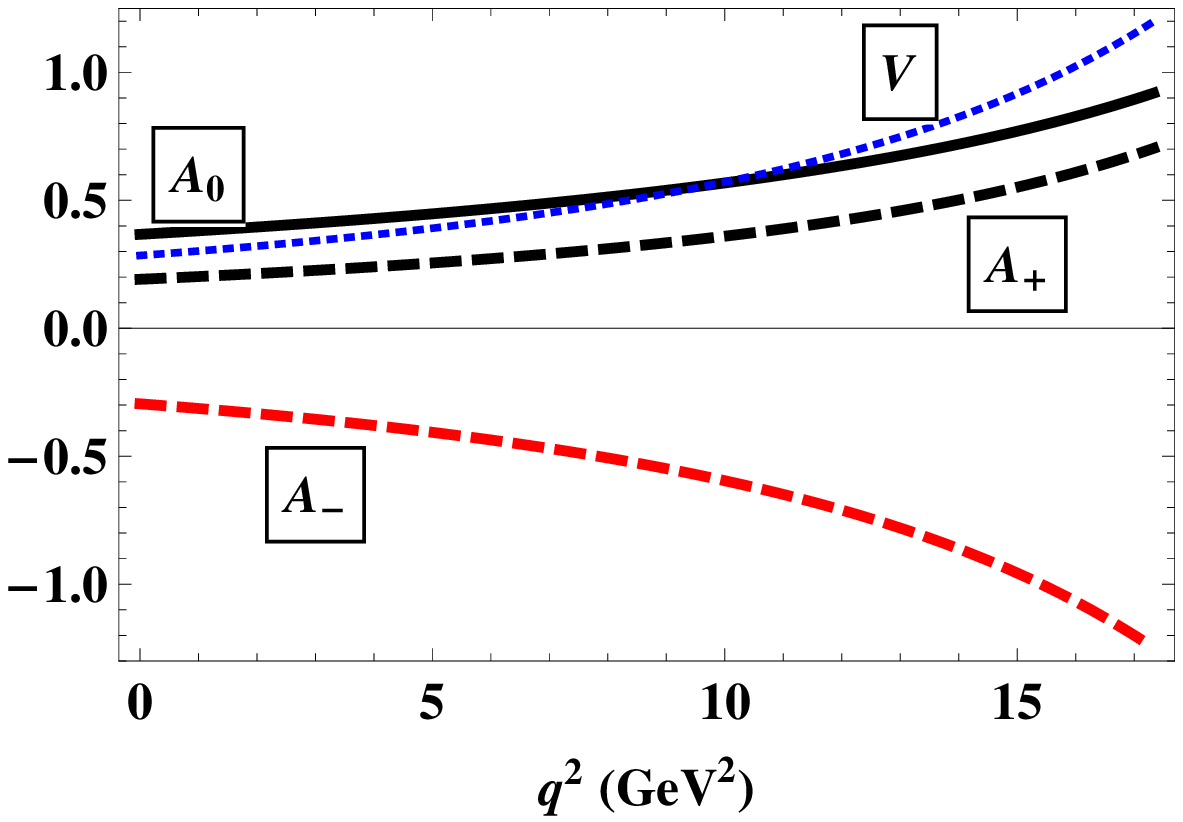} \\
\includegraphics[width=0.40\textwidth]{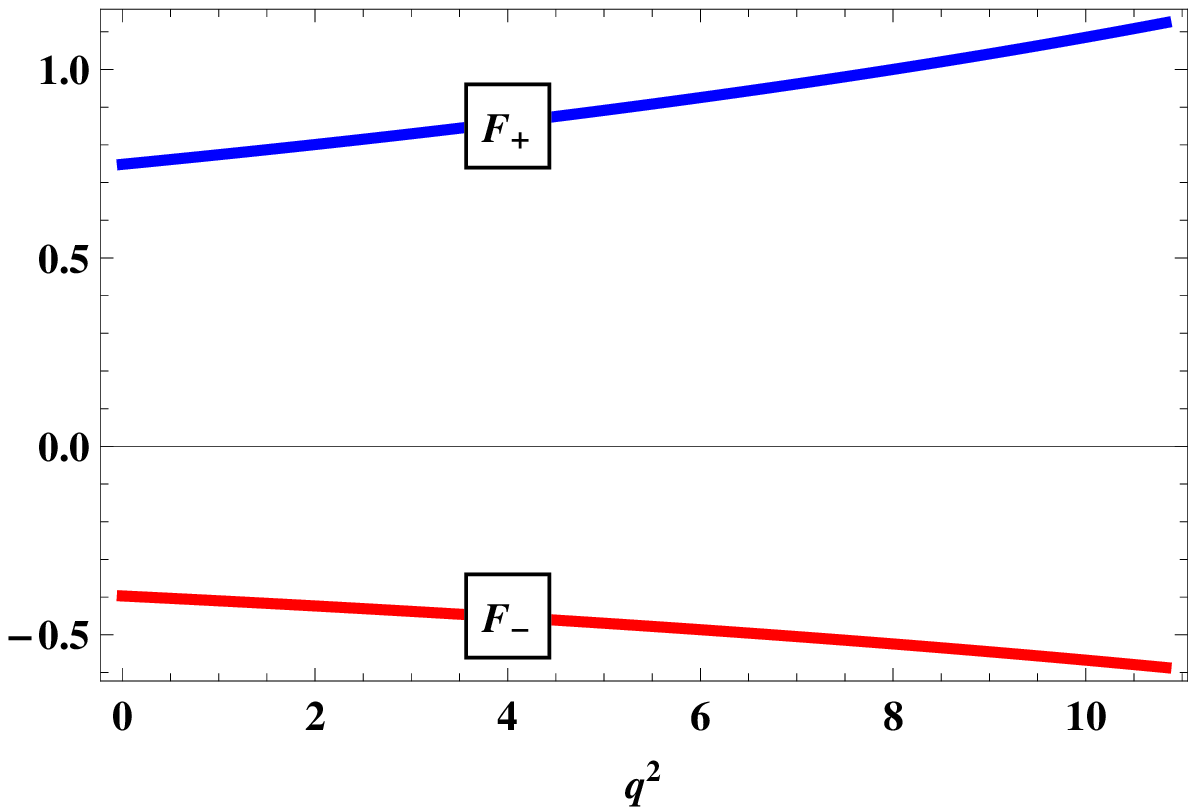} &
\includegraphics[width=0.40\textwidth]{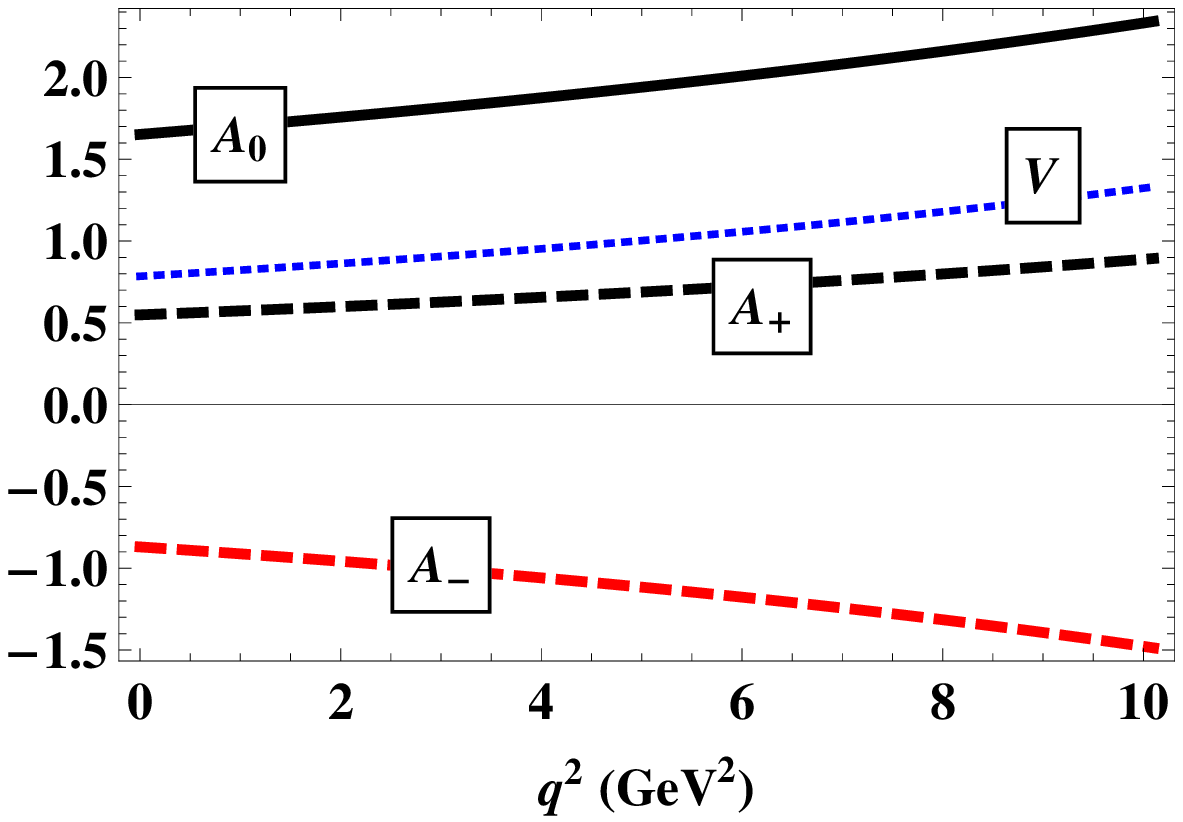} \\
\end{tabular}
\end{center}
\caption{\label{fig:FF}
Left panel: the form factors $F_+(q^2)$ and $F_-(q^2)$  for 
$B_c \to D,D_s,\eta_c$ transitions (from top to bottom).
Right panel: the form factors $A_0 , A_- , A_+ $ and $V$ for
$B_c \to D^\ast,D_s^\ast,\Jpsi$ transitions (from top to bottom).
}
\end{figure}

The form factors are calculated in the full kinematical region of 
momentum transfer squared. The curves are depicted  in Fig.~\ref{fig:FF}. 

The values of the form factors at maximum recoil ($q^2=0$) are given
in Table~\ref{tab:FF}.
\begin{table}[b]
\caption{$q^2=0$ results for the various form factors.}
%\vspace{0.3cm}
\begin{center}
\def\arraystretch{1.3}
\begin{tabular}{|c|c|c|c|}
\hline
  & $B_c\to D$ & $B_c^{+}\to D_s $  & $B_c\to \eta_c $  \\
\hline\hline
$F_{+}(0)$ &   0.186  &   0.254            &  0.74  \\ 
\hline
$F_{-}(0)$ & $-0.160$  & $-0.202$	  & $-0.39$ \\ 
\hline\hline
         & $ B_c \to D^\ast $  & $ B_c \to D_s^\ast $ & $ B_c\to \Jpsi $  \\
\hline\hline
$A_{0}(0)$ &  0.276  	      &   0.365                 &  1.65  \\ 
\hline
$A_{+}(0)$ &  0.151  	      &   0.190                 &  0.55  \\ 
\hline
$A_{-}(0)$  & $-0.236$         &  $-0.293$               & $-0.87$ \\ 
\hline
$V(0)$      &   0.230          &   0.282                &   0.78\\ 
\hline
\end{tabular}
\label{tab:FF}
\end{center}
\end{table}

\clearpage

\section{Numerical results}
\label{sec:numerics}

We are aiming to compare our results with those obtained
by the ATLAS~\cite{Aad:2015eza} and LHCb~\cite{Aaij:2013gia}  Collaborations. 
They reported the results of measurements of the ratios of the branching
fractions:
\be
\R_{\Ds/\pi^+} = 
\frac{\Br_{\Bc\to\Jpsi\Ds}}{\Br_{\Bc\to\Jpsi\pi^+}},
\quad
\R_{\DsStar/\pi^+} = 
\frac{\Br_{\Bc\to\Jpsi\DsStar}}{\Br_{\Bc\to\Jpsi\pi^+}},
\quad
\R_{\DsStar/\Ds} = \frac{\Br_{\Bc\to\Jpsi\DsStar}}{\Br_{\Bc\to\Jpsi\Ds}}
\label{eq:ExptRat}
\en
and the transverse polarization fraction in $B_c^+ \to J/\psi D_s^{*+}$ decay 
which is determined to be 
\be
\frac{\Gamma_{++}}{\Gamma} = 
\frac{\Gamma_{++}(B_c^+ \to J/\psi D_s^{*+})}
{\Gamma(B_c^+ \to J/\psi D_s^{*+})}.
\label{eq:ExptPol}
\en

First, we show up the input parameters used in calculations.
The central values of the CKM-matrix elements are taken from 
the PDG~\cite{Olive:2016xmw} and shown in Table~\ref{tab:CKM}.
\bgroup 
\def\arraystretch{1.5}
\begin{table}[htbp]
\caption{Values of the CKM-matrix elements.}
\vspace*{2mm}  
\centering
 \begin{tabular}{cccccc}
\hline
$|V_{ud}|$ & $|V_{us}|$ & $|V_{cd}|$ & $|V_{cs}|$  & $|V_{cb}|$  & $|V_{ub}|$ \\
\hline
   0.974  &  0.225    &   0.220   &   0.995    & 0.0405      & 0.00409 \\
\hline
 \end{tabular}
\label{tab:CKM}
\end{table}
\egroup
The central values of the relevant meson masses are taken from 
the PDG~\cite{Olive:2016xmw} and shown in Table~\ref{tab:mass}.
\bgroup 
\def\arraystretch{1.5}
\begin{table}[htbp]
\caption{Values of meson masses in GeV.}
\vspace*{2mm}  
\centering
 \begin{tabular}{ccccccc}
\hline
$m_{B_c}$ & $m_{\eta_c}$ & $m_{J/\psi}$ & $m_D$ & $m_{D^\ast}$  & 
                                     $m_{D_s}$  & $m_{D_s^\ast}$ \\
\hline
 6.275  &  2.983   &   3.097   &   1.869   & 2.010 & 1.968 & 2.112 \\
\hline
 \end{tabular}
\label{tab:mass}
\end{table}
\egroup

In Table~\ref{tab:Bc-nonlep} we show the values of branching fractions
obtained in this work for two different set of the Wilson coefficients.
One can see the difference is almost a factor of two between them.
Note that the values obtained with old set  $a_1=1.14,\,\,a_2=-0.20$
are very close to the predictions given in our previous paper
\cite{Ivanov:2006ni}.

We also calculate the widths of the decays $B_c\to M_{c\bar c}\pi$
to be able to compare with available experimental data.
Their analytical expressions are given by 
\be
\Gamma(B_c\to \pi^+ M_{\bar cc})  = 
\frac{G^2_F}{16\,\pi}\frac{ |{\mathbf p_2}|}{m^2_1}
\left|V_{cb}V_{ud}^\dagger a_1 f_\pi m_\pi\right|^2
\left(H_t^{B_c\to M_{\bar cc}}(m^2_\pi)\right)^2,
\label{eq:Bc-Jpi}
\en
where $M_{c\bar c}= J/\psi$ or $\eta_c$.

\bgroup 
\def\arraystretch{1.2}
\begin{table}[htbp]
\caption{\label{tab:Bc-nonlep}
         Branching ratios (in $\%$)
         of  nonleptonic $B_c$ decays obtained in this work
for two different set of the Wilson coefficients.}
\vspace*{2mm}  
\centering
\begin{tabular}{|l|c|c|c|}
\hline
 Mode & $a_1=+0.93$ & $a_1=+1.14$   & 
\cite{Ivanov:2006ni}  \\[-1.2ex]
     & $a_2=-0.27$ & $a_2=-0.20$  & \\ 
\hline
 $B_c \to \eta_c D_s$      & 0.22 & 0.50  & 0.44 
\\
 $B_c \to \eta_c D_s^{\ast}$ & 0.22 & 0.42  & 0.37 
\\
 $B_c  \to J/\psi D_s$      & 0.10 & 0.22  & 0.34 
\\
 $B_c  \to J/\psi D_s^{\ast}$  & 0.41 & 0.78  & 0.97 
\\
\hline
 $B_c \to \eta_c D$         & 0.0073 & 0.016  & 0.019 
\\
 $B_c \to \eta_c D^{\ast}$   & 0.0098 & 0.019  & 0.019 
\\
 $B_c \to J/\psi D$        & 0.0035   & 0.0074  & 0.015
\\
 $B_c \to J/\psi D^{\ast}$  & 0.017   & 0.031  & 0.045 
\\
\hline
\end{tabular}
\end{table}
\egroup

However, the ratio of the branching fractions is insensitive 
to the choice of the Wilson coefficients:
\be
R_{D^{*+}_{s}/D^{+}_{s}} =
\frac{B(B^{+}_{c}\to J/\psi D^{*+}_{s})} {B(B^{+}_{c}\to J/\psi D^{+}_{s})}
=\left\{\begin{array}{lr}
          3.55 & (a_1=1.14,\,a_2=-0.20) \\[2ex]
          3.96 & (a_1=0.93,\, a_2=-0.27) \\
          \end{array}
  \right.
\label{eq:ratio}
\en

Finally, we compare our results with available experimental data
and the results obtained in other approaches. For this purpose,
we take the Table from the paper \cite{Aad:2015eza} and add
our numbers. 
\begin{table}[htbp]
\begin{center}
\caption{
Comparison of the results for the ratios of branching fractions
with  those of ATLAS and LHCb Coll., and theoretical predictions. 
The used abbreviations are:\\
CCQM=covariant confined quark model (this work),\\
RCQM = relativistic constituent quark model~\cite{Ivanov:2006ni},\\
QCD PM=QCD Potential Model~\cite{Colangelo:1999zn},\\
QCD SR=QCD Sum Rules~\cite{Kiselev:2002vz},\\
BSW RQM=Wirbel-Stech-Bauer Quark Model~\cite{Dhir:2008hh},\\ 
LFQM=Light Front Quark Model~\cite{Ke:2013yka},\\
pQCD= perturbative QCD~\cite{Rui:2014tpa},\\
RIQM=Relativistic Independent Quark Model.
\label{tab:comparison}
}
\vspace{1em}
\begin{tabular}{c c c c r}
\hline\hline
$\R_{\Ds/\pi^+}$  & $\R_{\DsStar/\pi^+}$  & $\R_{\DsStar/\Ds}$  & 
$\Gamma_{\pm\pm}/\Gamma$  & Ref. \\
\hline\\[-0.9em]
$3.8 \pm 1.2$ & $10.4 \pm 3.5$ & $2.8^{+1.2}_{-0.9}$ & $0.38 \pm 
0.24$ &
ATLAS~\cite{Aad:2015eza} \\[0.2em]
$2.90\pm 0.62$ & -- & $2.37\pm 0.57$ & $0.52 \pm 0.20$ &
LHCb~\cite{Aaij:2013gia} \\
\hline
$1.29 \pm 0.26$ & $5.09 \pm 1.02$ & $3.96\pm 0.80$ & 0.$46 \pm 0.09$ & CCQM \\
$2.0$ & $5.7$ & $2.9$ & -- & RCQM \cite{Ivanov:2006ni} \\
$2.6$ & $4.5$ & $1.7$ & -- & QCD PM \cite{Colangelo:1999zn} \\
$1.3$ & $5.2$ & $3.9$ & -- & QCD SR \cite{Kiselev:2002vz} \\

$2.2$ & -- & -- & -- & BSW RQM \cite{Dhir:2008hh} \\
$2.06\pm 0.86$ & -- & $3.01\pm 1.23$ & -- & LFQM \cite{Ke:2013yka} \\[0.2em]
$3.45^{+0.49}_{-0.17}$ & -- & $2.54^{+0.07}_{-0.21}$ & $0.48\pm 0.04$ & 
pQCD \cite{Rui:2014tpa} \\[0.2em]
-- & -- & -- & $0.410$ & RIQM \cite{Kar:2013fna} 
\\
\hline\hline
\end{tabular}
\end{center}
\end{table}

One can see that the results of our calculations for the ratios
$\R_{\DsStar/\Ds}$ and $\Gamma_{\pm\pm}/\Gamma$ are consistent 
with measurements and other approaches. The results for the ratios 
$\R_{\Ds/\pi^+}$ and $\R_{\DsStar/\pi^+}$
are smaller than the measured values but the discrepancies do not exceed
two standard deviations.

\clearpage
\section{Summary}
\label{sec:summary}

We performed the calculations of the $B_c$ meson nonleptonic
decays: $B_c^+\to J/\psi \pi^+$, $B_c^+\to M_{c\bar c} D^{(\ast\,+)}_q$
where $M_{c\bar c}=\Jpsi$ or $\eta_c$,  $D^{(\ast\,+)}_q= D^{\ast\,+}_q$
or  $D^+_q$, and $q=s,d$. 

We compared the obtained results for several ratios of branching
fractions with those measured by the ATLAS and LHCb Collaborations
and other theoretical approaches.

We found that our prediction for the ratios
$\R_{\DsStar/\Ds}$ and $\Gamma_{\pm\pm}/\Gamma$ are consistent 
with measurements and other approaches. 
The results for the ratios $\R_{\Ds/\pi^+}$ and $\R_{\DsStar/\pi^+}$
are smaller than the measured values but the discrepancies do not exceed
two standard deviations.

\section{Acknowledgments}
Authors S. Dubni\v{c}ka, A. Z. Dubni\v{c}kov\'{a} and A. Liptaj acknowledge the support from the Slovak Grant Agency
for Sciences VEGA, grant No. 2/0153/17 and from the Slovak Research and Development Agency APVV, grant No.
APVV-0463-12.

\end{document}